\documentclass[preprint,showpacs,preprintnumbers,amsmath,amssymb]{revtex4}
\begin{document}

\title{Evidence for non-linear quasiparticle tunneling between fractional
quantum Hall edges}

\author{Stefano Roddaro, Vittorio Pellegrini, and Fabio Beltram}
\affiliation{NEST-INFM and Scuola Normale Superiore, I-56126 Pisa, Italy}

\author{Giorgio Biasiol, Lucia Sorba\footnote{Also with: Dipartimento di Fisica, Universita' di 
Modena e Reggio Emilia, I-43100 Modena, Italy}}
\affiliation{NEST-INFM and Laboratorio Nazionale TASC-INFM, Area Science Park, I-34012 Trieste, Italy}

\author{Roberto Raimondi}
\affiliation{NEST-INFM and Dipartimento di Fisica "E. Amaldi", Universita' di Roma Tre, Via
della Vasca Navale 84, I-00146 Roma, Italy}

\author{Giovanni Vignale}
\affiliation{Department of Physics, University of Missouri, Columbia, Missouri
65211, USA}

\begin{abstract}
Remarkable nonlinearities in the differential tunneling conductance between
fractional quantum Hall edge states at a constriction are observed in the
weak-backscattering regime. In the $\nu $ = 1/3 state a peak develops as
temperature is increased and its width is determined by the fractional
charge. In the range $2/3 \le \nu \le 1/3$ this width displays a symmetric
behavior around $\nu $ = 1/2. We discuss the consistency of these results 
with available theoretical predictions for inter-edge quasiparticle 
tunneling in the weak-backscattering regime.
\end{abstract}
\maketitle

Under the application of intense perpendicular magnetic fields, the kinetic energy
spectrum  of the two dimensional electron gas (2DEG) becomes quantized and
breaks into a sequence of macroscopically degenerate Landau levels.
The properties of electron states in this regime are largely driven by
Coulomb interactions. At some  ``magic" fractional values of the filling
factor $\nu $, in particular, the 2DEG condenses into collective
phases: the fractional quantum Hall (FQH) effect is the most remarkable
fingerprint of such incompressible quantum fluids \cite{aron,prange}.
Laughlin showed \cite{laug} that the emergence of such collective states
implies the existence of new quasiparticles carrying fractional charge.
Additionally, at the strong magnetic fields characteristic of the FQH effect
the lowest-energy charged excitations are confined at the edge of the sample in
one-dimensional branches whose excitations are forced to propagate
only in one direction. Wen \cite{Wen1} was the first to describe such
one-dimensional modes in terms of a chiral Luttinger liquid.
Wen's proposal stimulated a considerable amount of work aimed at
understanding the properties of this non-Fermi liquid state \cite{kane}.

The departure from the Fermi-liquid behavior is predicted to show up in many
properties of an interacting one-dimensional system \cite{kane92}. Most of the
experiments in quantum Hall systems concentrated on the observation of
the power-law behavior of the electron tunneling from a metal to the FQH edge
\cite{chang1,chang2,Grayson} and between two FQH edges
\cite{Milliken,glattli} in the {\em strong backscattering} limit. Resonant
tunneling in this regime was also explored \cite{Grayson2}. The inter-edge
tunneling process, in particular, can be experimentally induced at a
quantum point contact (QPC) constriction. In the {\em strong
backscattering} regime one observes the tunneling of {\it electrons}
between two quantum Hall fluids separated by the QPC. For simple fractions
(i.e. $\nu = 1/q$, where $q$ is an odd integer), this leads to a dc tunneling
current  at temperature T=0 given by $I_{T} \propto V_T^{(2/\nu - 1)}$. Notably
$I_T$ vanishes when the bias voltage $V_T$, with $V_T$ labeling the potential difference
between the two edges, tends to zero \cite{kane}.

In the opposite limit of {\em weak backscattering} the quantum Hall fluid
is weakly perturbed by the QPC constriction.   
In this case the inter-edge tunneling current (again, at $\nu =1/q$) 
consists of Laughlin quasiparticles
of charge $e^*=\nu e$ that scatter between the edges through the quantum
Hall fluid (see Fig.1 panel $(a)$). At T=0  the quasiparticle tunneling
rate is predicted to grow at low voltages  as $I_T \propto V_T^{(2\nu -
1)}$ in contrast to the electron-tunneling case discussed above.
This remarkable nonlinear behavior is removed at finite temperature.  
When $V_T$ falls below a critical value
$V_{T,max}$ of the order of  $k_BT/e^*$,
the tunneling current reverts to the
linear ohmic behavior $I_T \propto V_T$.  In the differential tunneling
characteristics ($dI_{T}/dV_{T}$) this leads to a peak centered at $V_T=0$ with a width
$\Delta V_T \simeq 2V_{T,max}$. This phenomenology was first predicted by
Wen \cite{Wen} who showed, in particular, that the width of the
finite-temperature peak is related to the fractional charge of the
quasiparticle.
The complex phenomena associated to non-linear
quasiparticle tunneling, however, are still largely experimentally unexplored.
During recent years
magneto-transport experiments in the {\em weak backscattering} regime 
were reported and concentrated on shot noise measurements aimed at detecting
the fractional charge of the quasiparticle \cite{noise}.
\par
In this Letter we report the observation of non-linear inter-edge tunneling
in the {\em weak backscattering} regime. We present the
differential tunneling conductance as a function of bias voltage in a wide
range of temperatures and filling factors. In the FQH regime our data 
cannot be described by electron tunneling but display the features of Wen's theory 
of quasiparticle tunneling. We demonstrate that while the differential tunneling
characteristic shows the tendency towards a diverging behavior as the
temperature is lowered at $\nu = 1/3$, it develops a peak centered at $V_T$ = 0 
as the temperature is increased.
Width and shape of this zero-bias peak
are determined by the fractional charge of the quasiparticle, consistently with
Wen's predictions.
We also discuss the results obtained for filling factors between
$\nu $ = 2/3 and $\nu $ = 1/3 where the zero-bias peak is observed even at
the lowest probed
temperature of 30 mK. Finally, we show that the evolution of the width of the
differential conductance peak displays an unexpected symmetry  around $\nu
$ = 1/2 not explained by current theories. We believe that these results
combine to provide the first  evidence of non-linear quasiparticle
tunneling in FQH systems.

The devices here studied were realized starting from a high-mobility
GaAs/Al$_{0.3}$Ga$_{0.7}$As two-dimensional electron gas (2DEG) with
low-temperature
mobility $\mu \sim 1 \times 10^6\,{\rm cm^2/Vs}$ and two-dimensional
electron density $n \sim 5 \times 10^{10}\,{\rm cm^{-2}}$. The 2DEG was
located $140\,{\rm nm}$ from the surface. 
The measurements discussed below were preformed in a dilution
refrigerator. The QPC constriction was nanofabricated on Hall-bar mesas (width of 80 $\mu
m$) using e-beam lithography and Al metallization. The width and length of the
QPC constriction were $300\,{\rm nm}$ and $600\,{\rm nm}$, respectively.

Differential inter-edge tunneling conductance as a function
of bias voltage $V_T$ was controlled by exploiting the QPC (see
Fig.1 panel $(a)$) in order to force edge states to flow close to each
other in the constriction. In the measurements presented here the
split-gate was biased at $V_g = -0.4\,{\rm V}$ which is just below the 2DEG depletion leading to 
2D-1D threshold at zero magnetic field. At these bias conditions therefore the conduction takes place through
the point contact only. This ensures that edge states indeed propagate 
inside the QPC while they are still separated enough to avoid strong
interactions. A current $I$ with both dc and ac components was supplied to
the Hall bar. This current causes a  Hall voltage drop  $V_H =
\rho_{xy}I$  (with dc and ac components) between the counter-propagating
edge channels within the
constriction \cite{footnote1}. In the weak backscattering regime $V_H$
coincides with the tunneling bias i.e., $V_T=V_H$.
Four-wire differential-resistance measurements were carried out
using an ac lock-in technique (see Fig.1 panel $(a)$) with $0 \le
I_{dc}\le 45 $ nA and $I_{ac}$ = 250 pA.
The latter value was chosen in order to have an acceptable signal-to-noise
ratio without introducing unphysical nonlinearities in the tunneling
current. The quantity that is measured in our configuration is the 
resistivity drop at the constriction (the differential
longitudinal resistance $dV/dI$). In the weak backscattering regime $dV/dI$ is
directly related to the differential tunneling conductance by:
\begin{equation}
dV/dI=\rho_{xy}^2dI_T/dV_T~.
\label{dVdI}
\end{equation}

The presence of residual backscattering outside the constriction leads to a
background signal superimposed to (\ref{dVdI}).
In the present devices this background is significant at $\nu = 1/3$. Even away from the
constriction the longitudinal resistivity 
is around $4k\Omega $, i.e. about $30\% $ of the value measured at the
constriction in the bias range of interest for the weak backscattering regime. 
However it does not display a sizeable variation as a function of 
the tunneling bias (data not shown) at least for low values of the tunneling bias. 
This and additional control experiments carried out at lower values of $V_g$  allow us to
unambiguously attribute the observed structures in $dV/dI$ to quasiparticle
tunneling at the QPC.

Panel $(b)$ in Fig.1 shows the longitudinal resistance
$\rho_{xx}$  at $I_{dc}=0$ as a function of magnetic field at $V_g$ = -0.4 V. This
measurement displays Shubnikov-de-Haas
oscillations associated to the quantum Hall states in the presence of the
constriction. At $I_{dc} \neq 0$ marked nonlinearities are observed in the FQH regimes.
An example is reported in the main panel of Fig.1 where two representative differential conductance
curves at filling factors $2$ and $1/3$ are shown.
As expected, the behavior in the integer regime is linear even at
the lowest temperatures explored. In order to emphasize the differences observed in 
the fractional regime, the marked nonlinear behavior measured in the case 
$\nu $ =1/3 is shown at the comparatively high temperature of  T = 400mK.

Figure 2 (panel a) shows the temperature evolution at $\nu=1/3$ 
for temperatures up to 900 mK. At this FQH state, the lowest-temperature
curve (at 30 mK) shows a minimum at zero bias. The zero-bias peak, however, is recovered
at higher temperatures.
It develops above $400\,{\rm mK}$ and tends to disappear as temperature
is increased above $900\,{\rm mK}$.  We can understand the data in Fig.
2 in the framework of the {\em weak-backscattering} theory for inter-edge tunneling
originally  proposed  by Wen \cite{Wen}.  We recall that in the
weak-backscattering regime the constriction is a small perturbation for
edge-state propagation
and induces a limited and localized backscattering. This limit can be
quantified comparing the tunneling current ($I_T$) to the total current
($I$) flowing through the device. We define the scattering to be {\em
weak} when $I_T \ll I$ (this condition could also be stated
as $dV/dI \ll h/\nu e^2 \sim 75k\Omega$).
In this regime we expect,
\begin{eqnarray}\label{IT}
I_T\left(V_T\right)=\frac{2\pi\left| t\right|^2}{\Gamma(2g)}
\left|\frac{2\pi T}{T_0}\right|^{2g-1}\left |\Gamma \left
(g+\frac{ix}{2\pi}\right )\right|^2
% \times\\ \nonumber
%\times B\left(\nu+\frac{ix}{2\pi},\nu-\frac{ix}{2\pi}\right)
\sinh\left(\frac{x}{2}\right),
\label{IV}
\end{eqnarray}
%where $B(x,y)=\Gamma(x)\Gamma(y)/\Gamma(x+y)$
where  $g=e^{*2}/\nu e^2$, $x=e^*V_T/k_BT$, $\Gamma$ is the Euler Gamma-function, and $t$ is the
inter-edge tunneling amplitude.   Note that in Wen's original theory the
chiral Luttinger liquid was assumed to exist only at the filling factors of
the fractional quantum Hall effect.   In (\ref{IT}) however, the
filling factor $\nu$ is allowed to vary continuously.  Such an extension of
the original formulation  can be justified on the basis of a hydrodynamic
model, which allows us to derive a continuum of Luttinger liquids with
continuously varying $\nu$ \cite{Dagosta}.

The behavior of $I_T$ depends crucially on the
relative size of $e^*V_T$ and $k_BT$. At low temperatures (\ref{IT})
predicts the nonlinear behavior  $I_T \propto V_T^{2g-1}$, which, for $g <
1/2$, leads to a growing current with decreasing bias: this is the signature of the 
{\it overlap catastrophe}.  At higher temperatures (\ref{IT}) predicts  an ohmic
behavior $I_T \propto V_T$. The crossover between the two regimes occurs at $V_T=V_{T,max}$, 
where $I_T$ reaches a maximum.
Figure 2 (panel $b$) shows the
calculated differential tunneling conductance  (the derivative of
(\ref{IT})) at $\nu = 1/3$ ($g=1/3$) at different temperatures. The
peak at zero bias arises from the ohmic region of the $I_T-V_T$ relation.
As the temperature lowers this peak is  expected to grow in intensity without saturation 
and to shrink in size. Indeed the
width of this peak (defined here as the distance  between the two zeroes of
the differential conductance) is $2V_{T,max}\sim 4.79k_BT/e^*$ and is
directly related to the effective charge of the quasiparticle involved in
the tunneling process \cite{Wen}.  For $V_T>V_{T,max}$ the differential
conductance becomes negative, and eventually tends to zero as $
V_T^{2g-1}$.

Figure 2 (panel $c$) reports the experimental results obtained for
temperatures higher than $400\,{\rm mK}$. The tunneling conductance in this
range of temperatures is characterized by a zero bias enhancement and the
observed overall trend is in good  agreement with the calculated behavior
shown in Fig.2 (panel $b$) and discussed above.  The agreement is
particularly satisfactory for what concerns the width of the conductance
peak. At higher temperatures the peak broadens and its amplitude decreases
following qualitatively the theory. At temperatures higher than $1K$ the
tunneling is completely linear in this voltage range.  The data also show
evidence of a {\it negative} contribution to the tunneling conductance at larger
values of the bias voltage: again this is in qualitative agreement with the
theory although the large background rises the average value of the
conductance above zero.

At temperatures lower than $400\,{\rm mK}$ (see Fig.2, panel $a$), 
however, a crossover to a completely different behavior is
observed. The tunneling conductance exibits a minimum at zero bias. 
The disappearance of the peak can be related to its progressive shrinking 
with temperature: for a given current modulation intensity a threshold temperature
can be estimated below which the peak cannot be detected. In 
the weak backscattering limit $V_{T,ac} \sim  \rho_{xy}I_{ac} \sim 0.02 mV 
\sim 100\,{\rm mK}$. This value is not in agreement with our experimental finding 
and suggests that the system by lowering the temperature may evolve 
into the strong backscattering regime \cite{nota28}.
A further mechanism for the zero-bias suppression can be associated to negative interference  
between spatially separated tunneling events within the QPC constriction. This effect
gets stronger with decreasing temperature and can contribute to the suppression 
at zero bias \cite{chamon}. 
 
Next we examine the dependence of tunneling current on filling factor. 
Experimental results are summarized in the color plots of
Fig.3 for different temperatures (here higher values of the
tunneling conductance are in yellow, lower values in black).
The measured $I_T-V_T$ relation was linear at integral filling factor (data not shown) and
for temperatures above 900 mK.
Otherwise, above 400 mK, the zero-bias peak is visible in the whole range $1/3 \leq
\nu \leq 2/3$,
and presents an interesting evolution: the peak
is strongest and narrowest at $\nu=1/3$ and $\nu=2/3$ (the two most
prominent QHE fractions) and rapidly broadens and loses its strength as
$\nu =1/2$ is approached from either side.  Notably its width evolves symmetrically about $\nu=1/2$.
This is shown in more detail in Fig.4 where the differential tunneling
conductance is plotted at three representative values of the magnetic field. 
The inset of Fig.4 reports
the  width of the zero-bias peak as a function of magnetic
field at T=400 mK (filled circles) and T = 500 mK  (open circles). A
similar behavior was found also at T = 700 mK (data not shown).

At the moment, a convincing theory of this is not available.  
Equation \ref{IT} is not particle-hole
symmetric under any reasonable assumption.  Setting $g=\nu$ yields a
tunneling conductance curve with a peak at zero bias: however the width of
the peak increases monotonically with $\nu$,
while its amplitude decreases. It should be pointed out that, at $\nu =
2/3$ (and more generally at
filling factors of  the Jain sequence $np/(np+1)$), the result of (\ref{IT}) 
with $g=\nu$ is in agreement with the weak backscattering
theory of composite edge states developed by Kane and
Fisher \cite{aron}. Finally, we should like to point out that
also the width of the zero-bias peak observed at $\nu > 1/3$ and low
temperature is not compatible with (\ref{IT}). One is tempted to associate the non-monotonic behavior of the peak
with the different structure of the zero-temperature fixed points controlling the charge mode for the Jain sequences
with positive and negative $p$. However, additional theoretical analysis
of quasiparticle tunneling at non-quantized filling factor values is
needed \cite{footnote2}.

In conclusion, we reported a marked non-linear behavior in
the inter-edge tunneling in the fractional quantum Hall regime in the
weak-backscattering limit. The observation of a zero-bias peak in the
differential
tunneling conductance has been interpreted as evidence for quasiparticle
tunneling between fractional edge states.
In selected ranges of temperatures and filling factors our data are
consistent with Wen's
theory of quasiparticle tunneling. Results as a function of filling factors
reveal intriguing
feautures not predicted by currrent theories.

We acknowledge R. D'Agosta, M. Grayson and V. Piazza for useful discussions. 
This work was supported in part by INFM

\newpage

\begin{figure}
\caption{Main panel: differential tunneling conductance ($dI_{T}/dV_{T}$) 
as a function of driving current ($I_{dc}$) at
the constriction for filling factors $\nu=2$ at T = 30 mK 
and $\nu=1/3$ at T = 400 mK. Panel $(a)$: sketch of the device and experimental set-up. 
panel $(b)$: longitudinal resistivity $\rho_{xx}$ as a function of 
magnetic field at $I_{dc}$ = 0 and T = 30 mK.}
\end{figure}

\begin{figure}
\caption{Panel $(a)$: differential tunneling conductance ($dI_{T}/dV_{T}$) for
filling factor $\nu=1/3$ at different temperatures 
(30 mK, 100 mK, 200 mK, 300 mK, 400 mK, 500 mK, 700 mK, 900 mK from bottom to top).
Panel $(b)$: Calculated $dI_{T}/dV_{T}$ (derivative of Eq. 2) 
in the weak-backscattering regime at T = 500 mK, 700 mK and 900 mK.
Panel $(c)$: selected  differential tunneling conductance curves at the same temperatures of panel (b).}
\end{figure}

\begin{figure}
\caption{Color plots of the differential tunneling conductance ($dI_{T}/dV_{T}$) as a function of the driving current $I_{dc}$ 
and magnetic field ($V_T =\rho_{xy}(B)I_{dc}$) at different temperatures.
$\nu = 1/3$ occurs at $B\approx 6 T$. Bright yellow regions correspond to high value of the tunneling conductance.}
\end{figure}

\begin{figure}
\caption{Representative differential tunneling conductance ($dI_{T}/dV_{T}$) at three
values of magnetic field and T = 500 mK. The inset reports the evolution of the peak 
width (as derived from a Lorentzian best-fit procedure) 
as a function of magnetic field for T= 400 mK (filled circles) and T = 500 mK (open circles). 
(see Fig.1 panel $(b)$ to relate magnetic field to the filling factor.)}
\end{figure}


\begin{references}
\bibitem{footnote1}It can be shown, by an elementary application of
the  Landauer-B\"uttiker formalism that this relation remains valid in the
presence of backscattering.
\bibitem{nota28}
In the strong backscattering regime the amplitude of the tunneling
voltage modulation changes from 
$V_{T,ac} \sim  \rho_{xy}I_{ac} \sim 0.02mV \sim 100\,{\rm mK}$ 
to the much larger value  $V_{T,ac} = \rho_{xx}I_{ac}$,
which diverges as $T^{-4}$. When $V_{T,ac}$  becomes much larger than 
the width of the zero-bias peak $\sim k_BT/e^*$  the peak is effectively 
washed out in the measurements.
\end{references}
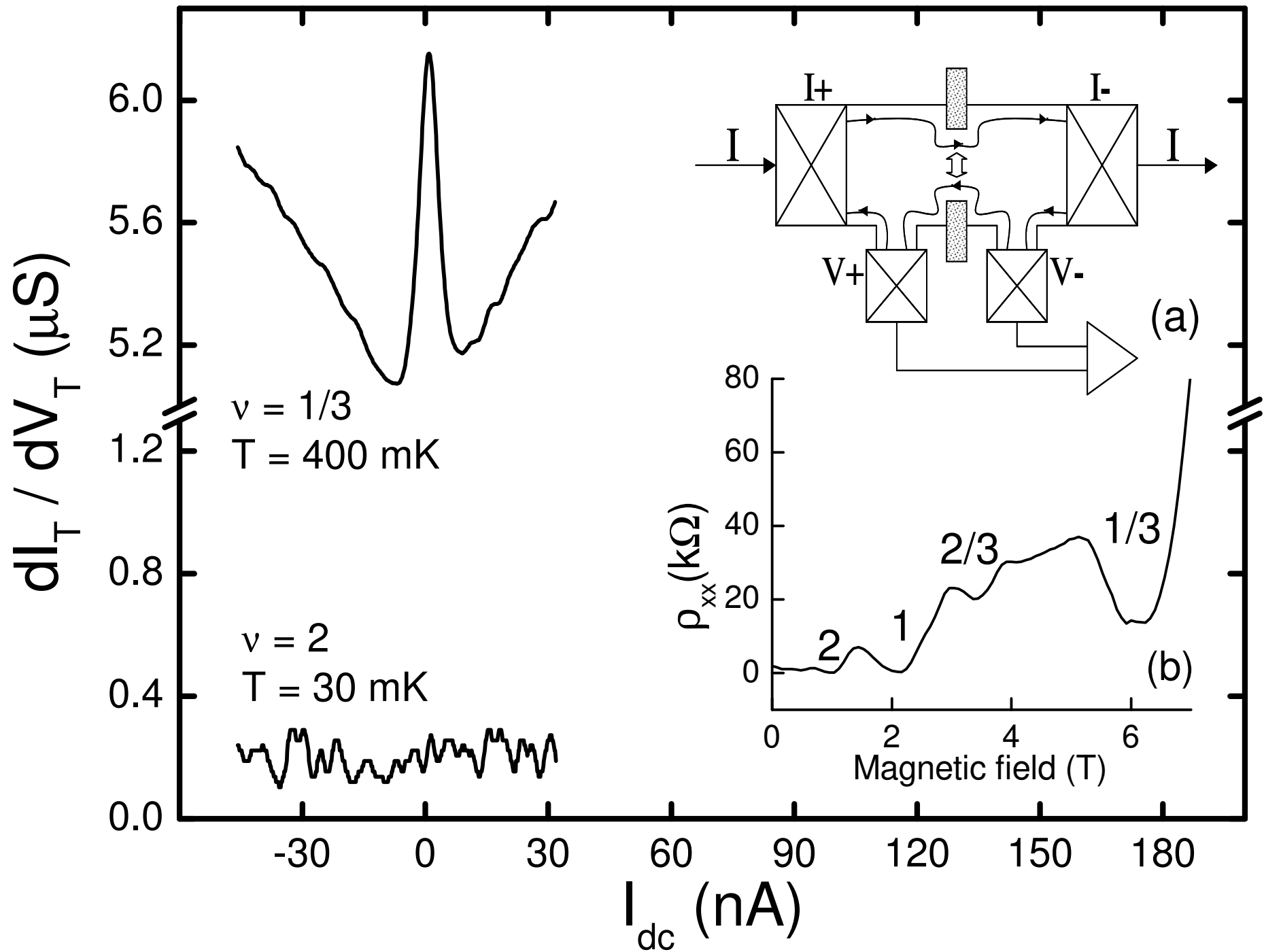
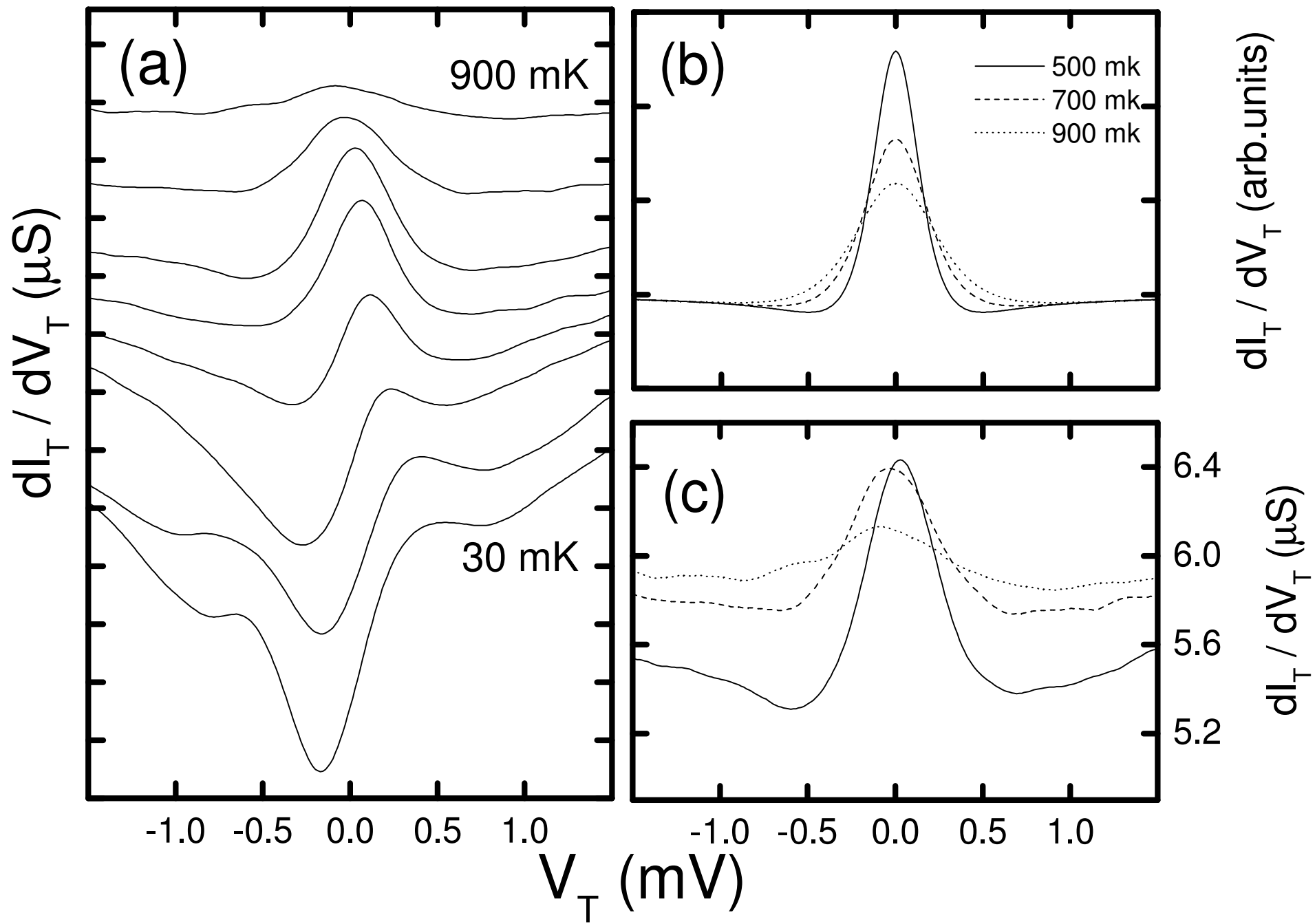
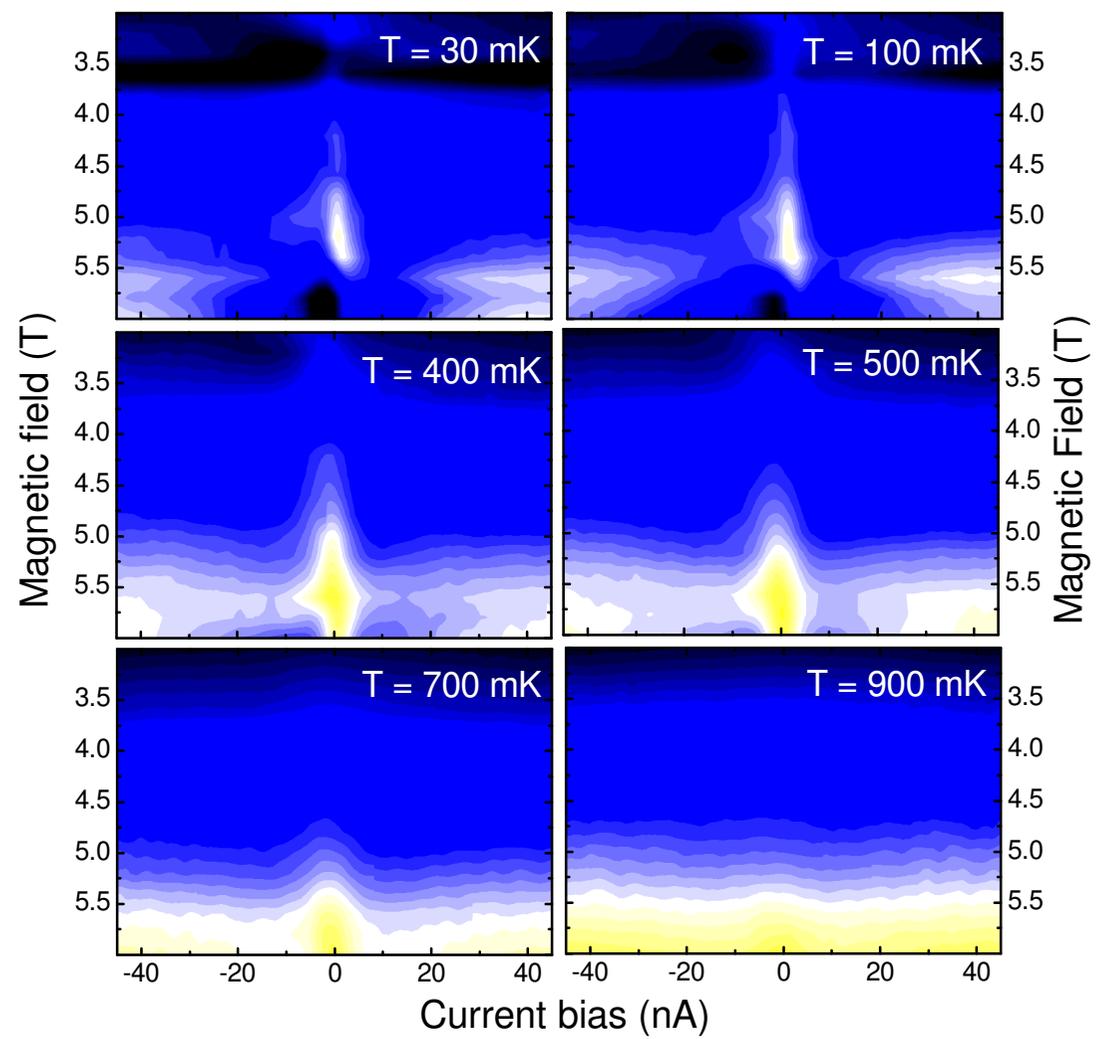
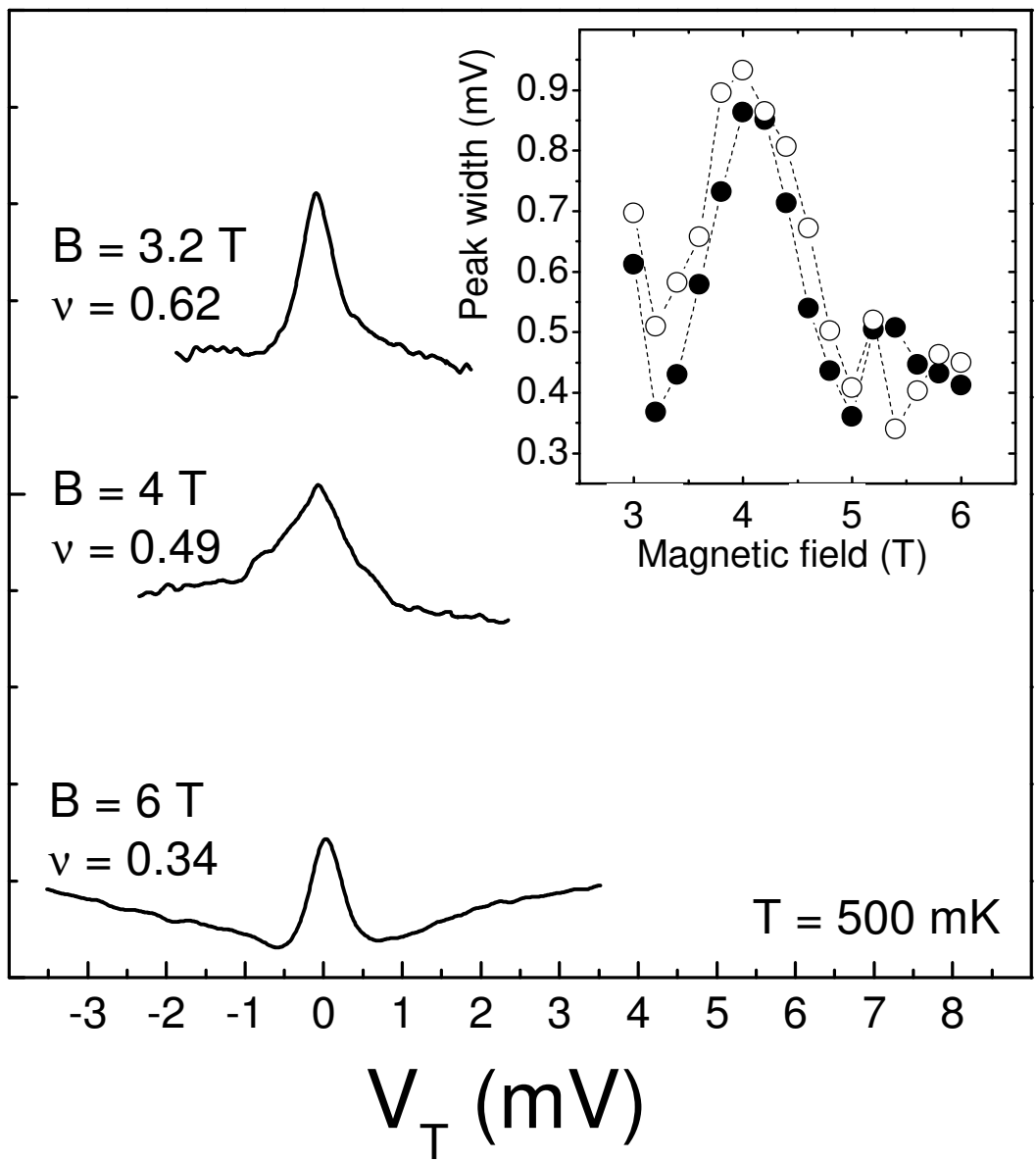
\end{document}